\def\year{2020}\relax
\documentclass[letterpaper]{article} 
\usepackage{aaai20}  
\usepackage{times}  
\usepackage{helvet} 
\usepackage{courier}  
\usepackage[hyphens]{url}  
\usepackage{graphicx} 
\usepackage{makecell}
\usepackage{graphicx}  
\usepackage{paralist}
\usepackage{enumitem}
\usepackage{threeparttable}
\usepackage{tablefootnote}

\usepackage{array}
\newcolumntype{P}[1]{>{\centering\arraybackslash}p{#1}}

\title{Exploring Effectiveness of Inter-Microtask Qualification Tests in Crowdsourcing}

\author{Masaya Morinaga\textsuperscript{\rm 1}, Susumu Saito\textsuperscript{\rm 1,2}, Teppei Nakano\textsuperscript{\rm 1,2}, Tetsunori Kobayashi\textsuperscript{\rm 1}, Tetsuji Ogawa\textsuperscript{\rm 1}\\
\textsuperscript{\rm 1}Department of Communications and Computer Engineering, Waseda University, Tokyo, Japan.\\
\textsuperscript{\rm 2}Intelligent Framework Lab, Tokyo, Japan.\\
\{morinaga, susumu, teppei, koba, ogawa\}@pcl.cs.waseda.ac.jp 
}
\begin{document}

\maketitle

\begin{abstract}

Qualification tests in crowdsourcing are often used to pre-filter workers by measuring their ability in executing microtasks.
While creating qualification tests for each task type is considered as a common and reasonable way, this study investigates into its worker-filtering performance when the same qualification test is used across multiple types of tasks.
On Amazon Mechanical Turk, we tested the annotation accuracy in six different cases where tasks consisted of two different difficulty levels, arising from the identical real-world domain: four combinatory cases in which the qualification test and the actual task were the same or different from each other, as well as two other cases where workers with Masters Qualification were asked to perform the actual task only.
The experimental results demonstrated the two following findings: 
\textit{i)} Workers that were assigned to a difficult qualification test scored better annotation accuracy regardless of the difficulty of the actual task; 
\textit{ii)} Workers with Masters Qualification scored better annotation accuracy on the low-difficulty task, but were not as accurate as those who passed a qualification test on the high-difficulty task.


\end{abstract}

\section{Extended Abstract}
To achieve accurate crowd-based data annotation, requesters are still facing challenges in selecting better techniques for pre-filtering crowd workers.
For example, filtering by workers' profiles such as language ability or task approval rate is known as one of the common methods~\cite{pee:vos:acq:2014}.
While this technique can be implemented by simple configuration, such rough statistics do not always directly lead to workers' actual task performance, and their performance in rejecting spammers and impatient workers is also limited.
Amazon Mechanical Turk (AMT) has a badge given to officially-certified workers, called Masters Qualification\footnote[1]{https://www.mturk.com/worker/help}, which can be used to filter workers by the presence of the badge.
However, this still remains a task-independent qualification with no clear criteria~\cite{kaplan2018}; in fact, some study reported that Masters Qualification was not effective~\cite{Rouse_2020}, which necessitates the use of the qualification to be further explored.


Qualification tests are known to be another reasonable technique for filtering workers based on their actual answers made on requesters' own exercise microtasks~\cite{heer_2010}.
Although it enables requesters to measure workers' task-dependent capabilities, it also has several trade-offs, such as it requires a fair amount of time in building microtasks and monetary cost in executing them.
A good workaround would be sharing the same qualification test among multiple microtask types, but to the best of our knowledge, there is no prior research that studied worker-filtering capabilities of such a practice.


In this study, the three following hypotheses were tested for developing a better practice in creating qualification tests:
1) The worker-filtering result becomes the most accurate when the qualification test has the same task as the production task;
2) the worker-filtering result can be diverted to other task types that are similar to that of the qualification test; and
3) at least for high-difficulty tasks, task qualifications have better worker-filtering performance than Masters Qualification.


\begin{figure}[tb]
\centering
\includegraphics[width=\linewidth]{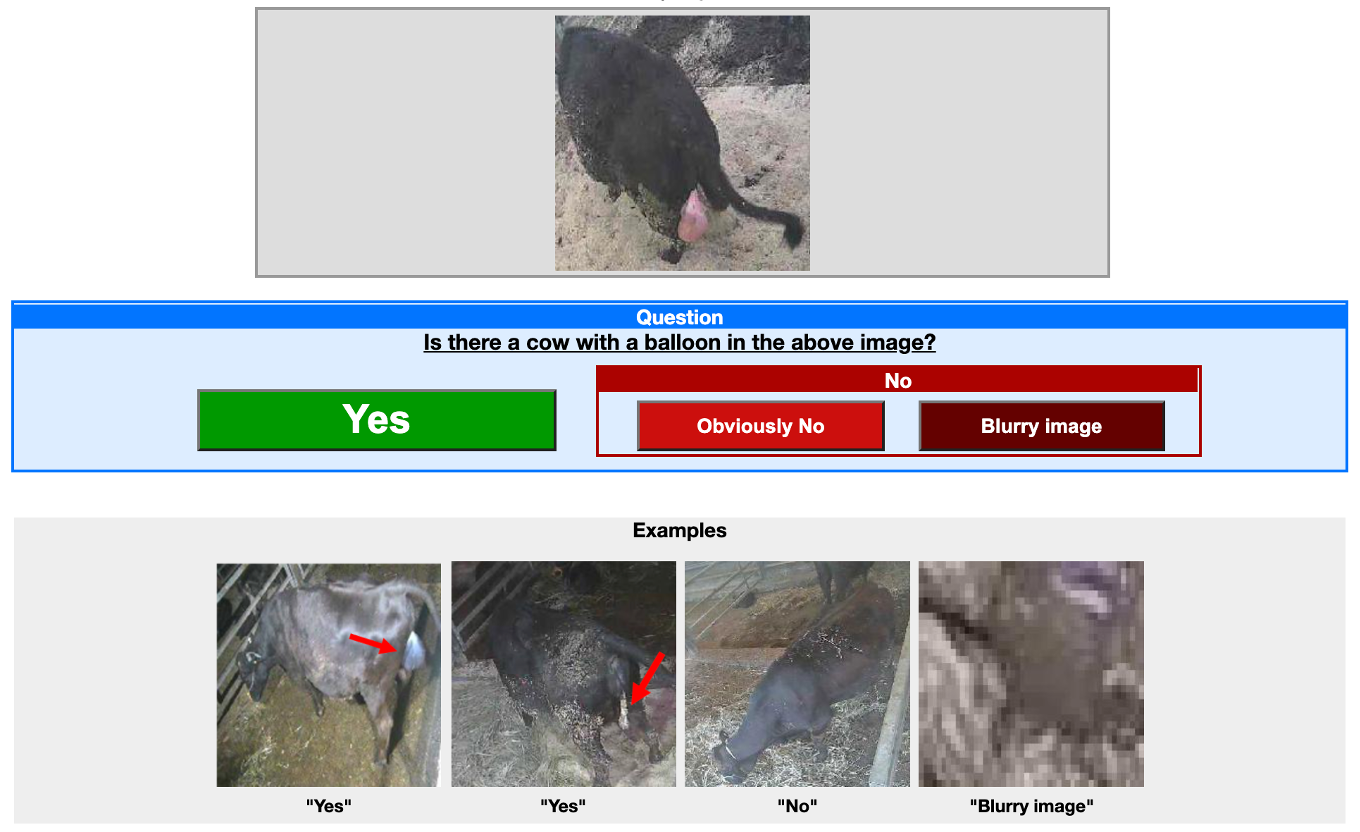}
\caption{Microtask UI for Balloon task}
\label{fig:balloon_screen}
\end{figure}

\begin{figure}[tb]
\begin{center}
\includegraphics[width=\linewidth]{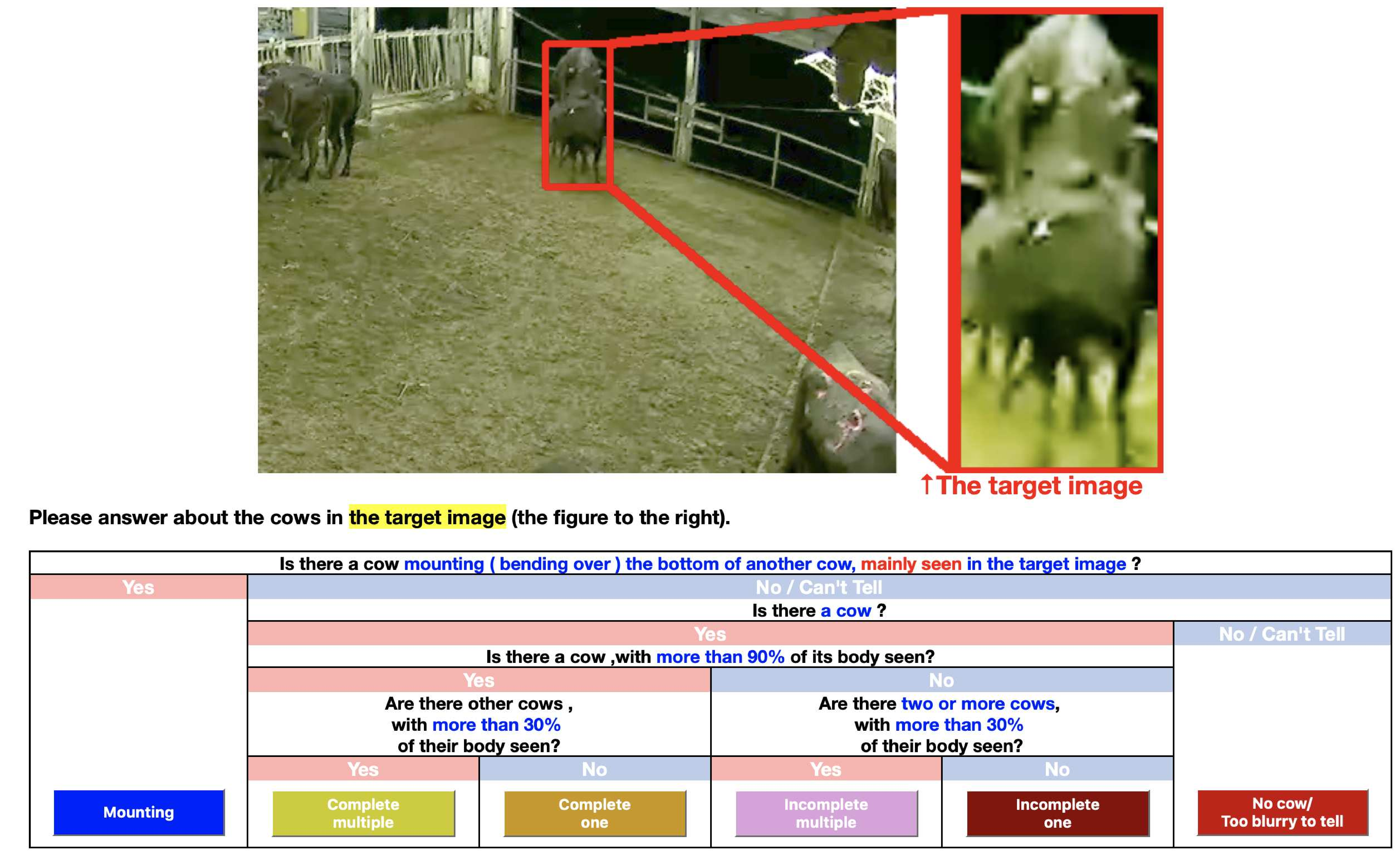}
\caption{Microtask UI for MT task}
\label{fig:mt_screen}
\end{center}
\end{figure}

To this end, we designed an experiment to investigate workers' answering accuracy in two types of annotation tasks based on a real-world domain, with two different levels of difficulty.
In this study, we picked annotation tasks on cattle images, captured by cameras installed on livestock farms~\cite{Hyodo2020}.
The first task is called the ``Balloon'' Task, and is a low-difficulty task.
In the task, workers were shown one of the region images detected by YOLOv2~\cite{redmon2016yolo9000} and asked to annotate whether or not a balloon-like object (\textit{i.e.,} allantochorion and fetal membrane) was seen around the cow's buttock (see Figure~\ref{fig:balloon_screen} for the microtask interface).
Ten images selected for the qualification test consisted of five images with ``Yes'' (\textit{positive}) labels and five with ``No'' (\textit{negative}) labels respectively, as their ground-truths.
Among the target subjects, only workers who answered 8 or more images correctly on the qualification test were sent an invitation link to the production task.
The second task is called the Cattle Mounting (``MT'') Task, and is a high-difficulty task.
Being shown a region image of two or fewer cows detected by YOLOv3~\cite{redmon2018yolov3}, workers were asked to judge whether one cow straddled another cow.
Only if not, workers were instead asked to choose one label out of four predefined options for false samples (``Complete one (CO)'', ``Complete multiple (CM)'', ``Incomplete one (IO)'', ``Incomplete multiple (IM)''), that were set based on the number of the cow(s) and their occlusion level (see Figure~\ref{fig:mt_screen} for the microtask interface).
For the qualification test, seven images -- three from the positive case and four from the negative case, each of which belonged to either one of the predefined labels -- were selected.
Among the target subjects, only workers who answered 5 or more images correctly on the qualification test were sent an invitation link to the production task.

The used image regions totalled 2,000 samples for the Balloon task and 1,085 samples for the MT task.
All ground truth labels were given manually by the authors for both tasks;
Balloon task consisted of 372 ``Yes'' and 1,628 ``No'' labels, and
MT task had 140 ``MT'' (positive), 116 ``CO'', 108 ``CM'', 256 ``IO'', and 465 ``IM'' labels, respectively.
Labels were collected from three workers for each image.


We compared the annotation results for two tasks with different conditions for worker filtering.
First, we recruited workers in groups with four different settings: w/o qualification (no worker filtering applied), w/ Balloon qualification, w/ MT qualification, and w/ Masters Qualification.
The answers were collected for each worker group, and the final labels were determined by aggregating answers with weighted majority voting based on worker ability~\cite{Dawid79maximumlikelihood}.


The annotation accuracy and the number of workers are shown in Table\ref{tab:exp_acc}. 
In addition, statistically significant differences were confirmed by the chi-square test for all combinations ($p<0.01$)
Results showed that both Balloon and MT tasks were given more accurate labels by all worker groups with a test-based qualification, compared to those by workers without qualifications.
Between the two test-based qualifications, MT task had even greater annotation accuracy improvement.
This implies that the worker-filtering performance by conducting qualification tests can be better when actual tasks are more difficult.

Reflecting the results by worker groups, the worker group with MT qualification scored the best in the both task types.
This implies that easy microtasks could test for higher-difficulty qualification (from the same domain) to get even better annotation accuracy.
This result therefore did not support our first hypothesis -- leaving a relatively surprising suggestion that qualification tests may also be measuring task-independent workers' abilities, such as how diligently they answer the questions, rather than task-specific abilities.

Also interestingly, the qualification test seemed even effective when the tasks were different between the qualification test and the actual annotation task; the both cases still scored better annotation accuracy than that by workers without qualifications.
This supports our second hypothesis, suggesting that the filtering results can be diverted to other tasks if task domains are similar.

Lastly, workers with Masters Qualification also had an interesting trend in their annotation accuracy.
While they scored better performance on Balloon task than workers with Balloon qualification, their answering accuracy on the MT task was the worst among all other worker groups.
This indicates that, at least in difficult microtasks that require some knowledge from a specific domain, Masters Qualification may not be enough to find high-performance workers -- thus supporting our third hypothesis.


\begin{table}[tb] 
\begin{threeparttable}
\fontsize{9.0pt}{10.0pt} \selectfont
\caption{Annotation accuracy (\# of participants). On each task type, all possible pairs between ``w/o qual.'' and ``w/ *** qual.'' were significantly different by $p<0.01$, based on the chi-square test.} 
\label{tab:exp_acc}
\centering
\begin{tabular}{P{9.5mm}|P{13mm}P{14mm}P{14mm}P{13mm}}\hline
& w/o qual. & \makecell{w/ Masters\\qual.} & \makecell{w/ Balloon\\qual.} & \makecell{w/ MT\\qual.} \\\hline
Balloon & \mbox{0.913 (194)} & \mbox{0.944 (24)} & \mbox{0.932 (26)} & \mbox{\textbf{0.964} (16)} \\
MT      & \mbox{0.645 (40)}  & \mbox{0.528 (26)} & \mbox{0.717 (17)} & \mbox{\textbf{0.761} (19)} \\ \hline
\end{tabular}

\end{threeparttable}
\end{table}


In this study, three hypotheses were tested to find good practices for creating qualification tests.
We believe that this paper demonstrated very important findings for future requesters in designing their task qualifications, as well as for researchers in establishing a new guideline for applying more precise and less-effort worker filtering techniques.


\section{Acknowledgments}
This study was supported by 9th Research Support Program of the Casio Science Promotion Foundation.

\bibliography{library}
\bibliographystyle{aaai}

\end{document}